\documentstyle[12pt,psfig]{article}
\def\reference{\parskip 0pt\par\noindent\hangindent 0.5 truecm}

\begin{document}
%
% Title
%
\title{Driftscan Surveys in the 21cm Line with the Arecibo and
Nan\c{c}ay Telescopes}
\author{F. H. Briggs\\Kapteyn Astronomical Institute\\fbriggs@astro.rug.nl \and 
  E. Sorar\\University of Pittsburgh\\ertu@phyast.pitt.edu \and
  R. C. Kraan-Korteweg\\ DAEC, Observatoire de Paris, Meudon\\kraan@gin.obspm.fr \and 
  W. van Driel \\ USN, Observatoire de Paris, Meudon\\vandriel@gin.obspm.fr }
\date{} % leave blank
\maketitle

\begin{abstract}
Driftscan methods are highly efficient, stable techniques
for conducting extragalactic surveys in the 21cm line of neutral hydrogen.
Holding the telescope still while the beam scans the sky at the sidereal
rate produces exceptionally stable spectral baselines, increased
stability for RFI signals, and excellent diagnostic information about system
performance.  Data can be processed naturally and efficiently by grouping
long sequences of spectra into an image format, thereby
allowing thousands of individual
spectra to be calibrated, inspected and manipulated as a single
data structure with standard tools that
already exist in astronomical software. The behavior of spectral standing waves
(multi-path effects)
can be appraised and  excised in this environment, making observations
possible while the Sun is up.  The method is illustrated with survey data
from Arecibo and Nan\c{c}ay.
\end{abstract}

\bigskip

\section{The technique}

The designs of the Arecibo and Nan\c{c}ay telescopes make them particularly
well suited to taking data in a driftscan mode. Both have large
collecting areas and are therefore sensitive survey instruments.
They obtain the large collecting area by having much of their
structure fixed to the ground.  When they are used to
track specific celestial coordinates, the on-axis gain changes and the
far out sidelobes move in unpredictable ways, causing spill-over on the
ground and RFI to be time variable. These instabilities increase the level
of systematic uncertainties, thus increasing the difficulty of detecting
weak signals.  Driftscan observations avoid these problems since all
components of the structures are fixed relative to the ground, thus
achieving the full sensitivity of the large reflecting areas. Similar
arguments apply to more conventional radio telescopes, when variation
in the spillover causes fluctuations in the spectral baselines and when RFI
entering the receiving system is modulated by gain variations
in the far sidelobes as the antenna tracks celestial sources.
This  report is based on experience obtained in extragalactic HI
surveys using the Arecibo and Nan\c{c}ay Telescopes.

Sorar (1994) and Briggs  conducted the 
Arecibo HI Strip Survey in the driftscan mode
in order to determine the HI-mass function for
nearby extragalactic objects by surveying long strips at constant
declination. The
observations covered approximately 6000 independent sightlines to a 
depth of 7500~km~s$^{-1}$, but since the strips were retraced on many days
in order to increase the integration time on each sightline, nearly a million
individual spectra had to be calibrated, regrouped and averaged.  The
details of the observational technique were developed by 
Sorar (1994), and the results and followup to 
the survey are summarized by Zwaan et al (1996, these proceedings).

\begin{figure}
\centerline{
}
\caption{Nan\c{c}ay Raw Data Image. Passband calibrated spectra have been loaded
into the image in time sequence increasing from right to left. There are two 
slightly overlapping, 6.4 MHz wide spectral bands, as marked at the
right border.  A trace of the continuum as a function of right ascension
is drawn under the image. }
\end{figure}

The method is also in use for surveys at Nan\c{c}ay. One project, being 
conducted by   Kraan-Korteweg, van Driel,
Binggeli, and Briggs, will test the completeness to a depth of
2300 km~s$^{-1}$ 
of deep optical catalogs 
of dwarfs and low surface brightness galaxies
in the CnV~I cloud (Binggeli et al 1990). 
A sample of the raw data from the HI survey are shown in Figure 1.
The spectral passband calibration for this data
has been performed by computing the average spectrum of the entire 3.5 hour dataset, and each spectrum was divided by the average spectrum as it was loaded in time sequence into the columns of
the image.  Passage of a telescope beam over a background continuum 
source is registered in the image as a dark band. The residual
Galactic HI emission causes the splotchy horizontal
band across the image around the HI rest frequency.

Once the data is loaded in image format, images become the units in
which the data is stored, manipulated and displayed. For example,
continuum subtraction, averaging of data from different days and
smoothing can
be accomplished using procedures in familiar astronomical image processing
packages.  Figure 2 shows the processed
spectra for a total of $\sim$2500 sightlines in three adjacent
declination strips. The figure results from approximately 82,000 individual
spectra. There are 11 detected extragalactic signals resulting from
9 separate galaxies, plus a number of interesting features associated
with Galactic HI.

A second project now in progress at Nan\c{c}ay will observe ${\sim}6$\% 
of the sky to a depth of 4500~km~s$^{-1}$ with noise level ${\sim}23$ mJy
(5$\sigma$) for velocity resolution 20 km~s$^{-1}$.  The survey is
well matched to detecting nearby examples of gas-rich systems such as 
HI~1225+01 (Giovanelli \& Haynes 1989, Chengalur, Giovanelli \& Haynes 1995) 
and the circum-galactic ring in Leo (Schneider 1989), as
well as detecting a sample of several hundred normal galaxies.

\begin{figure}
\centerline{
}
\caption{Processed Nan\c{c}ay images for three declinations.  
Spectra are loaded in horizontal lines in this image, with right ascension
increasing upwards. Overlap of the spectral sub-bands has been removed, and
labels on the horizontal axis indicate velocity in km~s$^{-1}$. Eleven
detected signals are marked.}
\end{figure}

\section{Spectral ``standing waves''}

Faint periodic
fluctuations in the background noise are visible in the
low velocity ranges of each strip in Figure 2.  These are commonly called
``standing waves'' by radio spectroscopists.  In these Nan\c{c}ay data, they
occur in the correlator quadrants that have strong Galactic HI emission, but
not in the higher velocity quadrants.  The standing waves arise because
Galactic HI signal enters the receiving system by two paths of different
length.  In an ideal telescope this would not happen, but it is  
common   in the present design of radio telescopes, since 
radiation entering the radio receiver may have been weakly scattered by 
structure  that crosses the telescope's
aperture; this additional scattering can deflect off-axis 
radiation into the beam of the telescope. When two copies of a signal
take different paths to the receiver, 
the signal taking the longer path suffers a delay, and when the
autocorrelation function is computed in the spectrometer, a correlation
spike is
obtained in the delay channel corresponding to the extra path length. In
the Fourier transformation of the ACF to obtain the power spectrum, a
single-channel spike transforms to a sinusoidal variation across the band.
Thus, the path delay corresponds directly to the number of cycles of
standing wave across the spectral band.  For the high velocity ranges of
Figure 2, there is no interfering signal entering the system with a
time delay 
and with sufficient strength to generate standing waves, and therefore
the noise
characteristics in this redshift range are better behaved.

\begin{figure}
\centerline{
}
\caption{Schematic of multipath scattering at the Arecibo Telescope.
Representative off-axis rays, $R1$, $R2$, and $R3$, are seen traversing
paths of different length before arriving at the Arecibo line feed $F$. 
In this diagram, $R1$ takes a direct path to the feed, while $R2$ and $R3$
are scattered from different points ($S2$ and $S3$) 
on the support structure. Radiation
is actually scattered in many directions 
at the scattering points $S2$ and $S3$, 
but those rays that are scattered downward parallel to the incoming
on-axis rays are directed to be optimally reflected from the main
reflector directly toward the feed.
}
\end{figure}

Any broadband 
signal entering the receiving system in multiple copies with different
delays can give rise to these multipath effects. 
A strong  source of such radiation 
is the Sun during daytime observations, but terrestrially generated broadband
RFI, Galactic emission, strong radio continuum sources and
spillover can also do it.
The standing waves are often dominated by a single periodicity, representing,
for example, a round trip delay between the feed cabin and the dish
surface.  However,
a further complication is that antenna structures are often sufficiently
complex that several scatterering locations can contribute
(as illustrated for the Arecibo Telescope in Figure 3), 
producing signals
in several delay channels of the ACF.  When the Fourier transform is computed,
the spectral passband produced by the combination of multiple
sinusoidal components can be very complicated. Figure 4 illustrates the
complexity of standing wave patterns caused by the Sun 
in some Arecibo observations made during a late afternoon observation.
Note that  while the feed was
positioned at the same antenna coordinates, this same pattern repeated 
on successive (solar) days.  
There was a slow change in the pattern after several days
as the Sun moved in declination. 

\begin{figure}
\centerline{
}
\caption{Arecibo daytime standing waves. 
{\it Left.} The data image for a single circular polarization
after the standard calibration and continuum removal.
The image contains 1250 seven second time steps for a duration of
${\sim}$2.4 hours. The bright, spatially resolved 
galaxy M94 can be seen midway through the image in the low velocity
range. There is a faint linear feature running vertically in the image
at slightly lower velocity -- this is Galactic HI emission that is aliased
into the spectrum through the baseband filters.
{\it Right.} Harmonic content of each row in the image for first
63 harmonics. The amplitudes have been averaged by 4 time steps in this
image.
 }
\end{figure}

As well as varying in amplitude, standing waves can drift
in phase with time, causing them to fail to subtract exactly when the simpler
forms of passband calibration are applied.  This is  true of the
processing that has been done to obtain Figure 2, and
substantial residuals remain in the low velocity range.  There is a
time range in the observations shown in Figure 4 where the
drift is so rapid that the wave moves by a full turn of phase in just a few
minutes.
Fortunately, there is a straightforward way to
tackle this problem and remove even these complicated patterns from 
the driftscan data in an unbiased way. It begins 
by performing the Fourier transform of the power spectrum
$S_{t_i}(\nu)$ 
to obtain the amplitudes and phases of the standing wave components
during each time step $t_i$.
These harmonics can be written as complex coefficients 
${\bf{A}}_{n,t_i}= A_{n,t_i}\exp(i\phi_{n,t_i})$,
where $n$ labels the harmonic by 
the number of full periods of the wave across the
spectral band and $t_i$ indicates that the coefficients are expected to
change with time, as the spectra are recorded in discrete time steps.
In principle, the Fourier analysis of a string of real numbers, such as
these spectra, produces one half as many complex harmonic coefficients 
as there were spectral channels in the 
power spectrum, but, in practice, the standing waves 
arise from significant signal only at low values of $n$.
Thus, the standing wave content of an image-formatted database can
be described by another ``image'' (or table)
of complex numbers, with the same number
of time steps, but  many fewer values for ${\bf A}_n$ than are
required for the number of channels in the spectrum $S(\nu)$.

Figure 4 shows the time behavior of the harmonic content of the
first 63 harmonics in the Arecibo data image in the left part of the figure.
The standing waves appear with a variety of periodicities, with
several long-lived bursts in the harmonics around $n=20$, which corresponds to 
differential delays ${\sim}1 \mu$sec, the round-trip light travel
time between the Arecibo feed support structure
and the surface of the reflector.
There is substantial variability depending on where the scatterer is
located on the support structure.  
There are also some bursts of signal in the lowest
harmonics, possibly due to differential delays
between scattered paths such as $R3$ and $R2$ in
Figure 3, which are both scattered downward into the dish from different
points on the support structure;  alternatively, there could be scattering
from point $S2$ directly in the direction of the feed $F$, which would
also cause a fairly short differential 
delay and thus a long period standing wave.
The harmonic signature of a bright galaxy can also
be seen near the midpoint of the data set.  The faint, periodic, horizontal
striping is a result of the slightly variable correlator dump time, causing the
record integration time to beat with the 7 second grid spacing to
which the data was interpolated. 

Phase drift of an $n$ harmonic standing wave is described by watching the phase
term $\phi_n(t)$ from ${\bf{A}}_{n,t_i}$ vary with time. 
Figure 5 shows an example of the amplitude and phase data for a single 
harmonic $n=15$. The time variation of both
$A_n(t)$ and $\phi_n(t)$ can be efficiently tracked in time by 
sliding a window of 16 to 128 time steps along the table of ${\bf{A}}_{n,t_i}$
and then taking the Fourier transform of the complex time series 
of each harmonic within the windowed region.

A recipe for tracking and modeling a standing wave is summarized as follows:

\noindent (1) Compute the table (or image) of complex harmonic coefficients
 ${\bf{A}}_{n,t_i}$

\noindent (2) Separate the complex coefficients 
of the $n$th harmonic as a function
of time into a single long vector (such as the data plotted in Figure 5). 

\noindent (3) Subdivide the vector into short enough time
spans that the rate of drift is nearly constant over that time window.
The choice of  length for the time window depends on
how rapidly the rate of drift changes, since the window must be short enough
to track changes in the drift rate but also long enough to be immune to
signals that are localized in a single beam.  Of course, a choice
for the window length of $2^N$ time steps (with $N$ equal to an 
integer in the range 4 to 7) 
helps to increase the efficiency in computing the transform.
  
\noindent (4) 
The phase interpolation needed to track the waves is simplified if there
is overlap of the windows, and therefore the window was advanced
by either $2^{N-1}$ or $2^{N-2}$ time steps before 
recomputing the transform. 
Thus, a mathematical summary is: each window contains 
$p=2^N$ points taken from ${\bf{A}}_{n,t_i}$ with $t_i$ running
from $t_m$ to $t_{m+p-1}$.  The Fourier transform of this series of $p$
numbers produces $p$ complex coefficients 
${\bf{a}}_{n,{\delta}_k} = a_{n,{\delta}_k}\exp{(i\theta_{n,{\delta}_k})}$.
Here $a_{n,{\delta}_k}$ is the strength of the component drifting at
rate $\delta_k$ with phase $\theta_{n,{\delta}_k}$ at time $t_w=t_{m+p/2}$.
In principle,   
there may be well be a range of different standing wave drift rates $\delta$ 
contributing at any given time, 
since many locations on the support structure are capable of scattering.
In practice, for each window, we tabulated a  
$\delta_S$ and a $\theta_{n,{\delta}_S}$ corresponding to the 
$\delta_k$ and $\theta_{n,{\delta}_k}$ of the
strongest $a_{n,{\delta}_k}$ 
component in each time window, after 
checking for significance relative to the noise level.

\noindent (5) Depending of the degree of overlap of the windows, each time
step $t_i$ falls in either 2 or 4 windows.  Each window that produced 
valid measurements for $\delta_S$ and $\theta_{n,{\delta}_S}$ can be used to
form an estimate of the standing wave phase at $t_i$.  Thus,
a reasonable method for tracking the phase for the $n$th harmonic is
to compute an estimate $<\phi_{n,{t_i}}>$ for the phase at each time 
from the weighted vector average of the overlapping windows:
$$\eta e^{(i<\phi_{n,{t_i}}>)} = \left[\sum W_{n,w} (|t_i - t_w|)^\rho e^{(i(\delta_{S,w}(t_i - t_w)+\theta_{n,{\delta}_S,w})}\right]/\sum W_{n,w} 
(|t_i - t_w|)^\rho $$
The sums are taken over the 2 (or 4) windows
that overlap at $t_i$. The weighting factors
include $W_{n,w}$, which indicates the statistical significance of
the solution for the window $w$,  and a factor $(|t_i - t_w|)^\rho$,
 which gradually
transfers the weight among the windows by assigning the greatest emphasis to
the solution whose window is centered closest to $t_i$. The factor $\rho$ is
one of many possible weight adjustments.  An amplitude $\eta$ also results
from the vector average; $\eta$ is close to unity when there is close
agreement between the phase determined by all the windows included in the
average.

\begin{figure}
\centerline{
\psfig{figure=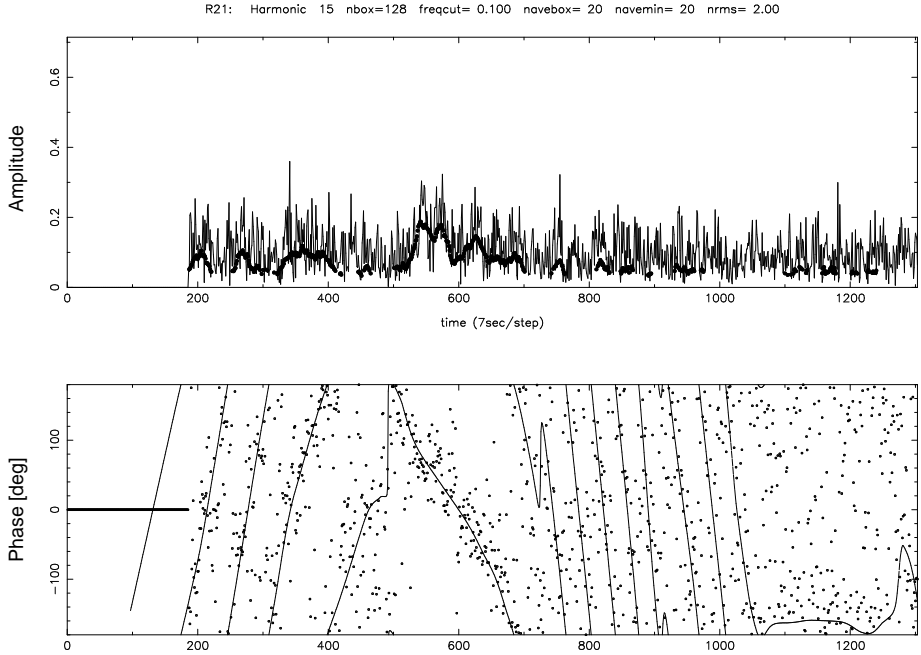,width=13cm}
}
\caption{Amplitude and phase for the $n=15$ harmonic standing wave.
{\it Top panel.}  Coefficient amplitudes plotted as a function of time
step (light noisy curve). 
The heavy solid points are vector averages of the amplitude (for
20 time steps), computed after application of the phase tracking
algorithm. Heavy points are plotted only when the amplitude surpasses
a set threshold.
{\it Bottom panel.}  Standing wave phases (points).  Smooth interpolated
curve results from the phase tracking algorithm. The curve is lighter
in regions where the signal to noise ratio for the wave is low.
}
\end{figure}

\noindent (6) Once a solution for phase tracking has been performed
(such as shown in the lower panel of Figure 5), a model 
for the temporal behavior of the wave amplitudes
 can be made by smoothing the time sequence of
$A_{n,t_i}\exp(i(\phi_{n,t_i}-<\phi_{n,{t_i}}>))$.  When the resulting wave
amplitudes surpass a set 
threshold for significance, they can be stored in a table of 
harmonic coefficients that can subsequently be used to generate models 
for the standing waves 
as a function of time (as shown in Figure 6) and then correct 
the observations 
by subtracting the standing wave model from the data image.
Note that this technique succeeds at doing little damage to the celestial
signals, since galaxies typically fall in only one beam and the fits are
derived from many beams.  This approach is far superior to simply ``nulling''
the ${\bf A}_n$ component, since this would throw away genuine information from the sky that is specific to each beam.

\begin{figure}
\centerline{
}
\caption{Standing wave edits.
{\it  Left panels.} Versions of the data image in Figure 4, but with
standing wave models subtracted. The data images are presented on the
same grayscale wedge as Figure 4.
{\it Right panels.} Standing wave models built from harmonic analysis 
of the data image in Figure 4, followed by application of a phase 
tracking algorithm  and reconstruction of a noise-free images of the standing waves. The upper model includes harmonics $n=$ 4 through 25, when they
are deemed significant. The lower model includes $n=$ 1 through 25. The
grayscale is about $3\times$ more sensitive for the models than for the
data images.
}
\end{figure}

While our experience at Arecibo and Nancay has shown that the
strongest standing waves are due to the sun, we have also seen
similar, but weaker, 
standing waves at night at Arecibo when observing in the vicinity of the
strong radio source
Taurus A (the Crab Nebula); in this case, the pattern repeated at the
same sidereal time day after day.  Very strong standing waves have also 
been generated by faulty equipment that generated intermittent, broadband
noise; these waves had fixed phase in successive
scans, since the source of the rfi
was fixed to the Earth and the antenna was stationary during the observation.

\section{Final Comments}

The driftscan technique for spectral line surveying offers a number of 
advantages over pointed observations: (1) Data quality is maximized since
the telescope is still relative to the ground; 
spillover, far sidelobes and on-axis gain
are constant. (2) System stability can be monitored with high precision.
(3) The observations are 100 percent efficient since data can be taken
continuously. (4) The telescope scheduling is simple, and observing staff
has little real-time responsibility, 
since the telescope is sitting still with
the brakes on. (5) Spectral passband fluctuations due to standing waves
(or any other instability that varies on time scales longer than a few
minutes) can be tracked and removed, permitting observations to be
made during the daytime. (6) The data is naturally processed in image-format;
the analysis is very simple and efficient, and allows large projects to be 
tackled using existing
software, which is proven and familiar.

\section*{Acknowledgments}

We are grateful to the telescope staffs at Arecibo and Nan\c{c}ay for their
assistance with the observations.  
The Nan\c{c}ay radio observatory is operated by the Observatoire
de Paris and associated as USR B704 to the French Centre
National de Recherche Scientifique (CNRS), with financial
support of the Region Centre. The Arecibo Observatory is part of the National
Astronomy and Ionosphere Center, which is operated by Cornell University under
agreement with the U.S. National Science Foundation.
The research by RCKK is being supported with an EC-grant.
The Arecibo HI Strip Survey received extensive support through 
NSF Grant AST~91-19930.

\medskip
\reference  Binggeli, B., Tarenghi, M, \& Sandage, A. 1990, A\&A, 228, 42
\reference  Briggs, F.H. 1996, this volume
\reference  Chengalur, J.N., Giovanelli, R., \& Haynes, M.P. 1995, AJ, 109, 2415
\reference  Giovanelli, R., \& Haynes, M.P. 1989, Ap.J., 346, L5
\reference  Schneider, S.E. 1989, Ap.J., 343, 94
\reference  Sorar, E. 1994, Ph.D. Thesis, University of Pittsburgh
\reference  Zwaan, M., Sprayberry, D., \& Briggs, F.H. 1996, this
volume.
\end{document}